\documentclass[aps,prx,reprint,preprintnumbers,superscriptaddress,nofootinbib,longbibliography,floatfix]{revtex4-2}
\pdfoutput=1
\usepackage{rotating}
\usepackage{array}
\usepackage{amsmath}
\usepackage[normalem]{ulem}
\usepackage{slashed}
\usepackage{booktabs}
\usepackage[pdftex,table]{xcolor}
\usepackage{units}
\usepackage{xfrac}
\usepackage{mathtools}
\usepackage{empheq}
\usepackage[]{units}
\usepackage{multirow}
\usepackage{amssymb}
\usepackage{url}
\usepackage{comment}
\usepackage{physics}
\usepackage{color,soul}
\usepackage{bbm}
\usepackage[caption=false]{subfig}

\usepackage{hyperref}
\hypersetup{
  colorlinks=true,
  citecolor=blue,
  linkcolor=blue,
  urlcolor=blue
}

\begin{document}

\title{Anomaly Detection under Coordinate Transformations}


\author{Gregor Kasieczka}
\email{gregor.kasieczka@uni-hamburg.de}
\affiliation{Institut f\"{u}r Experimentalphysik, Universit\"{a}t Hamburg, 22761 Hamburg, Germany}

\author{Radha Mastandrea}
\email{rmastand@berkeley.edu}
\affiliation{Department of Physics, University of California, Berkeley, CA 94720, USA}
\affiliation{Physics Division, Lawrence Berkeley National Laboratory, Berkeley, CA 94720, USA}

\author{Vinicius Mikuni}
\email{vmikuni@lbl.gov}
\affiliation{National Energy Research Scientific Computing Center, Berkeley Lab, Berkeley, CA 94720, USA}

\author{Benjamin Nachman}
\email{bpnachman@lbl.gov}
\affiliation{Physics Division, Lawrence Berkeley National Laboratory, Berkeley, CA 94720, USA}
\affiliation{Berkeley Institute for Data Science, University of California, Berkeley, CA 94720, USA}

\author{Mariel Pettee}
\email{mpettee@lbl.gov}
\affiliation{Physics Division, Lawrence Berkeley National Laboratory, Berkeley, CA 94720, USA}

\author{David Shih}
\email{dshih@physics.rutgers.edu}
\affiliation{New High Energy Theory Center, Rutgers University, Piscataway, NJ 08854, USA}

\begin{abstract}
    There is a growing need for machine learning-based anomaly detection strategies to broaden the search for Beyond-the-Standard-Model (BSM) physics at the Large Hadron Collider (LHC) and elsewhere.  %
    The first step of any anomaly detection approach is to specify observables and then use them to decide on a set of anomalous events.  One common choice is to select events that have low probability density.  It is a well-known fact that probability densities are not invariant under coordinate transformations, so the sensitivity can depend on the initial choice of coordinates.  The broader machine learning community has recently connected coordinate sensitivity with anomaly detection and our goal is to bring awareness of this issue to the growing high energy physics literature on anomaly detection. 
    In addition to analytical explanations, we provide numerical examples from simple random variables and from the LHC Olympics Dataset that show how using probability density as an anomaly score can lead to events being classified as anomalous or not depending on the coordinate frame.
\end{abstract}

\maketitle


\section{Introduction}
\label{sec:intro}

Given the vast parameter space of Beyond-the-Standard-Model (BSM) physics as well as the lack of recent discoveries, there has been a growing interest in new search techniques that reduce model dependence. A number of novel approaches based on machine learning (ML) have been proposed that can automatically identify anomalous regions of phase space~\cite{DAgnolo:2018cun,Collins:2018epr,Collins:2019jip,DAgnolo:2019vbw,Farina:2018fyg,Heimel:2018mkt,Roy:2019jae,Cerri:2018anq,Blance:2019ibf,Hajer:2018kqm,DeSimone:2018efk,Mullin:2019mmh,1809.02977,Dillon:2019cqt,Andreassen:2020nkr,Nachman:2020lpy,Aguilar-Saavedra:2017rzt,Romao:2019dvs,Romao:2020ojy,knapp2020adversarially,collaboration2020dijet,1797846,1800445,Amram:2020ykb,Cheng:2020dal,Khosa:2020qrz,Thaprasop:2020mzp,Alexander:2020mbx,aguilarsaavedra2020mass,1815227,pol2020anomaly,Mikuni:2020qds,vanBeekveld:2020txa,Park:2020pak,Faroughy:2020gas,Stein:2020rou,Kasieczka:2021xcg,Chakravarti:2021svb,Batson:2021agz,Blance:2021gcs,Bortolato:2021zic,Collins:2021nxn,Dillon:2021nxw,Finke:2021sdf,Shih:2021kbt,Atkinson:2021nlt,Kahn:2021drv,Aarrestad:2021oeb,Dorigo:2021iyy,Caron:2021wmq,Govorkova:2021hqu,Kasieczka:2021tew,Volkovich:2021txe,Govorkova:2021utb,Hallin:2021wme,Ostdiek:2021bem,Fraser:2021lxm,Jawahar:2021vyu,Herrero-Garcia:2021goa,Aguilar-Saavedra:2021utu,Tombs:2021wae,Lester:2021aks,Mikuni:2021nwn,Chekanov:2021pus,dAgnolo:2021aun,Canelli:2021aps,Ngairangbam:2021yma,Bradshaw:2022qev,Aguilar-Saavedra:2022ejy,Alvi:2022fkk,Dillon:2022tmm,Birman:2022xzu,Raine:2022hht,Letizia:2022xbe,Fanelli:2022xwl,Verheyen:2022tov,Caron:2022wrw,Dillon:2022mkq,Buss:2022lxw}.  These techniques can be categorized by their BSM hypotheses, which directly relates to their ML strategy.  While most proposals do not make these hypotheses explicit, the existing approaches typically posit one of three possibilities: (i) the BSM is rare: $p_B(x)$ is small for background probability density $p_B$ and for BSM data point $x$; (ii) the BSM is overdense: $p_S(x)/p_B(x)$ is large for signal probability density $p_S$; (iii) the BSM is more similar to known BSM models than to the background.  

These three possibilities approximately map onto unsupervised, weakly supervised, and semisupervised machine learning methods, respectively.  All three of these generic hypotheses are much weaker than the usual, fully supervised case where the hypothesis is very specific and involves assuming particular couplings, decay chains, masses, etc.  The unsupervised methods learn implicitly or explicitly the probability density $p_B$, and then anomalous events are defined by those with a low density, usually through the use of an anomaly score that is a proxy for $p_B$ (such as the loss function of an autoencoder - see Sec.~\ref{sec:unsup}).  Weakly supervised methods learn likelihood ratios between a target dataset and a reference (mostly anomaly-free) dataset.  Weakly supervised learning refers to learning with noisy labels -- in this context, the target dataset has a noisy label of `signal' while the reference dataset has a noisy label of `background'.  Semisupervised methods use a number of simulated signal models, often combined with (mostly anomaly-free) data.  This categorization is not unique and the names used here are based on their meaning in the high energy physics (HEP) ML literature (see Ref.~\cite{Karagiorgi:2021ngt} for a recent review). Most papers on anomaly detection for HEP fall in the unsupervised category, although the only existing ML-based anomaly detection physics results use weakly supervised learning~\cite{collaboration2020dijet,Shih:2021kbt}. For this paper, we consider only the unsupervised and weakly supervised cases, as they are the most commonly studied for HEP analyses.

Due to the ill-posed nature of attempting to identify \textit{any} anomaly, there is no one method that will be more sensitive than all others\footnote{In the limit of infinite statistics, flawless background estimation, and arbitrarily powerful ML model architecture/training, weakly supervised methods can be universally optimal.  Of course, this is never true in practice; see Ref.~\cite{Nachman:2020lpy}, App.~A.}.  Each approach has merits and disadvantages.  For example, previous works have explored the tradeoffs between unsupervised and weakly supervised learning~\cite{Amram:2020ykb,Collins:2021nxn,Kasieczka:2021xcg}.  In particular, Ref.~\cite{Collins:2021nxn} pointed out that in the context of resonance searches, weakly supervised methods may outperform unsupervised methods for relatively higher signal fractions because they can explicitly use the presence of the anomalies to guide their performance.  In contrast, unsupervised approaches are nearly independent of the presence of signal and so can maintain performance even at low signal fraction.  However, if the signal is in the bulk of the background distribution, then unsupervised methods may be unable to find it no matter how much signal is present.

Another core feature of anomaly detection approaches is their response to coordinate transformations. While likelihood ratios are independent of invertible coordinate transformations, the notion of an event being `rare' is inherently coordinate-dependent.  Since unsupervised methods cannot be guided by the presence of anomalies as in weakly supervised approaches, the selection of observables used for anomaly detection may be more important for unsupervised methods compared with weakly supervised approaches.  The fact that probability densities are not invariant under coordinate transformations is well known and the connection to anomaly detection has recently been explored in the broader machine learning community~\cite{e23121690}. Our goal is to bring awareness of this issue to HEP, where there are a growing number of proposals that make use of coordinate-dependent methods. While coordinate sensitivity is relevant for both achieving signal sensitivity and estimating the Standard Model background, we focus entirely on the former as it is usually the focus of recent anomaly detection proposals.

This paper is organized as follows.  Section~\ref{sec:landscape} provides a taxonomy of ML-based anomaly detection methods.  The statistical properties of coordinate transformations of observables are described in Sec.~\ref{sec:stats}.  Illustrative numerical examples are given in Sec.~\ref{sec:examples}, first with a simple, analytic example and then a more realistic example based on a dijet search at the Large Hadron Collider (LHC).  The paper ends with conclusions in Sec.~\ref{sec:conclusions}.

\section{Landscape of Anomaly Detection Methods}
\label{sec:landscape}

In this section, we provide a brief summary of unsupervised and weakly supervised anomaly detection methods. We also provide references to recent applications of these methods in the HEP field. 

\subsection{Unsupervised}
\label{sec:unsup}

One of the most popular approaches studied in the phenomenology literature is the autoencoder (AE).  The first AE approaches~\cite{Hajer:2018kqm,Heimel:2018mkt,Farina:2018fyg} worked by simultaneously training two neural networks: an encoder network $f:\mathbb{R}^N\rightarrow\mathbb{R}^M$ and then a decoder network $g:\mathbb{R}^M\rightarrow \mathbb{R}^N$.  The typical loss function is the mean squared error\footnote{Capital letters represent random variables and lower case letters represent realizations of the random variables.}: $\langle (g(f(X))-X)^2\rangle$.  For arbitrarily flexible networks and training procedures, $f\circ g$ could approach the identity.  To ensure this does not happen, the network capacities and training procedure are restricted and $M\ll N$. Anomalies are then characterised by high reconstruction loss $(g(f(x))-x)^2$ compared to the background. 

As with any compression algorithm, the autoencoder will maximize its efficiency if it dedicates its limited capacity based on the probability density of a given event.  For this reason, the AE implicitly\footnote{Vanilla AEs have a strong dependence on the ML architecture and training procedure, which means that they may not be as precise at estimating the density as other approaches.} estimates $p_B(x)$. Anomaly scores based on autoencoders can also be created to take advantage of the compressed latent space created by the algorithm. Those are often based on Variational Autoencoders (VAEs)~\cite{kingma2014autoencoding,Kingma2019} or similar methods, trained to generate a latent space with useful statistical properties~\cite{Cerri:2018anq,pol2020anomaly,Mikuni:2020qds,Jawahar:2021vyu,Cheng:2020dal,Bortolato:2021zic,Dillon:2021nxw,Jawahar:2021vyu,NAE,Dillon:2022mkq}. %

Beyond VAEs, other deep generative models proposed for unsupervised anomaly detection include Generative Adversarial Networks (GANs)~\cite{Goodfellow:2014:GAN:2969033.2969125,Creswell2018} and Normalizing Flows~\cite{10.5555/3045118.3045281,Kobyzev2020}. In all of these cases, the generative model is implicitly (GANs and VAEs) or explicitly (Normalizing Flows (NFs)) learning $p_B(x)$, so anomaly scores are directly linked to the probability density.  
In the case of GANs, anomalies can be identified by combining the generative model with an autoencoder \cite{knapp2020adversarially} and assigning an anomaly score to the reconstruction loss between the inputs and the generated outputs.  With a direct estimate of the density, the output of a NF can be used directly as an anomaly score~\cite{Verheyen:2022tov,Aarrestad:2021oeb,Buss:2022lxw}\footnote{Normalizing Flows have also been proposed for use as weakly supervised anomaly detection methods - see Ref.~\cite{Nachman:2020lpy,Hallin:2021wme,Butter:2022lkf,Raine:2022hht}.}.  A detailed comparison of various generative models on benchmark BSM signals was studied in Ref.~\cite{Aarrestad:2021oeb}.

\subsection{Weakly Supervised}

In contrast to unsupervised methods, weakly supervised approaches require two datasets: a reference and a target.  Some approaches  emphasize the estimation of the reference sample~\cite{Collins:2018epr,Collins:2019jip,Andreassen:2020nkr,Nachman:2020lpy,Hallin:2021wme,Hallin:2021wme,Raine:2022hht} and some approaches take the reference sample as given~\cite{DAgnolo:2018cun,DAgnolo:2019vbw,dAgnolo:2021aun,Chakravarti:2021svb,Letizia:2022xbe,2203.09601,Alvi:2022fkk}. Strategies for determining the reference sample span a spectrum ranging from signal-model agnostic and background-model dependent approaches using simulations to resonance searches where sideband information can be directly used to estimate a background-only reference. Hybrid methods have also been proposed, as in the case of creating noisy labels for weak supervision using unsupervised autoencoders~\cite{Amram:2020ykb}.

Once the reference sample is acquired, most methods estimate the likelihood ratio directly by training a classifier to distinguish examples from the target and reference datasets.  It is well-known that the output of a classifier trained with a standard loss function like binary cross-entropy is monotonically related to the likelihood ratio (see e.g. Ref.~\cite{hastie01statisticallearning,sugiyama_suzuki_kanamori_2012}).  Directly estimating probability densities and taking ratios has also been explored~\cite{Nachman:2020lpy}.

\section{Statistics of Coordinate Transformations}
\label{sec:stats}

In this section, we will review some elementary facts about probability densities and their applications to anomaly detection\footnote{Note that methods that do not exactly learn the density like vanilla autoencoders may have additional susceptibilities to variable transformations.}.

\begin{figure}[h!]
    \centering
    \subfloat[]{\includegraphics[width=0.45\textwidth]{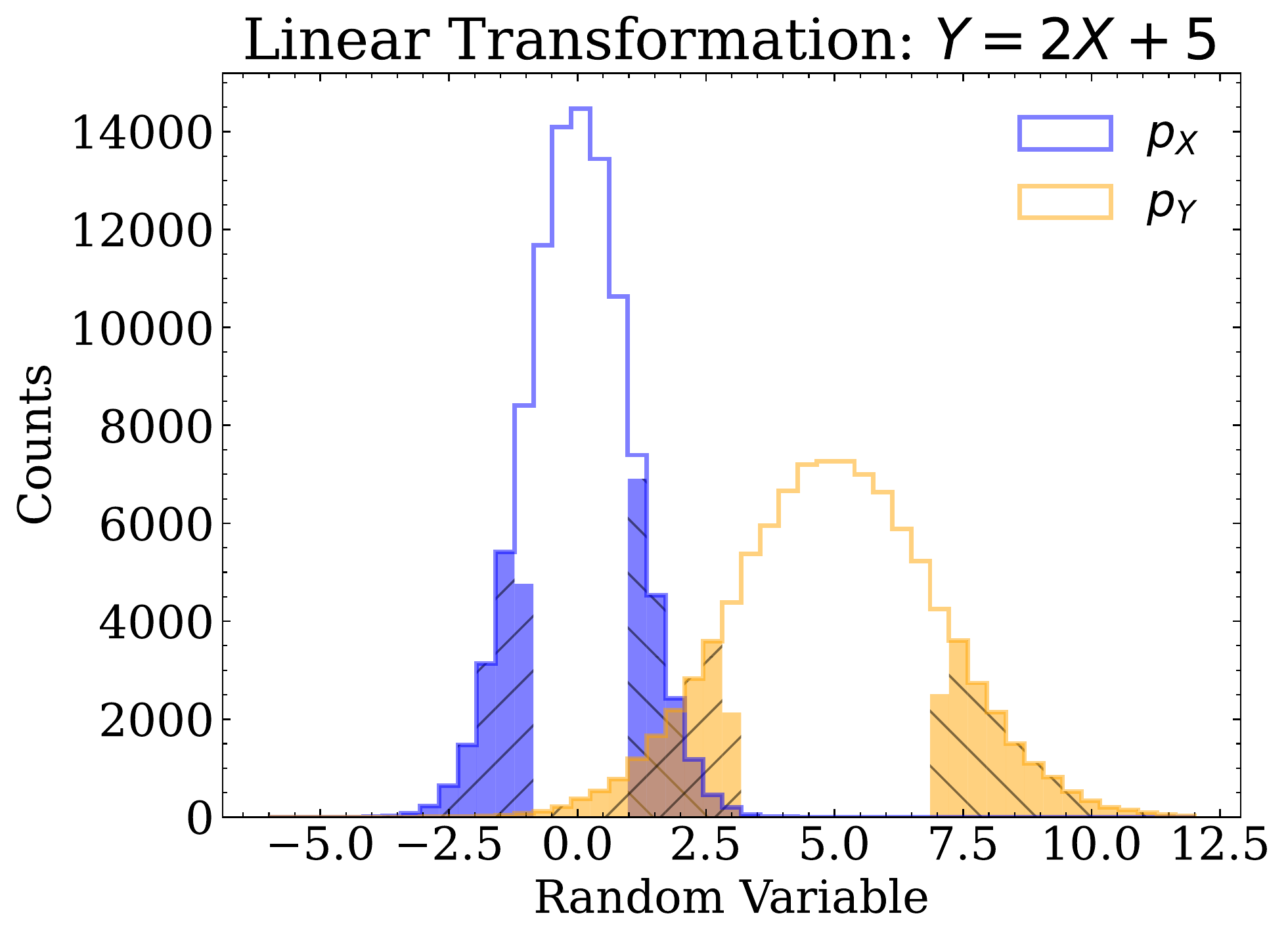}}\\
    \subfloat[]{\includegraphics[width=0.45\textwidth]{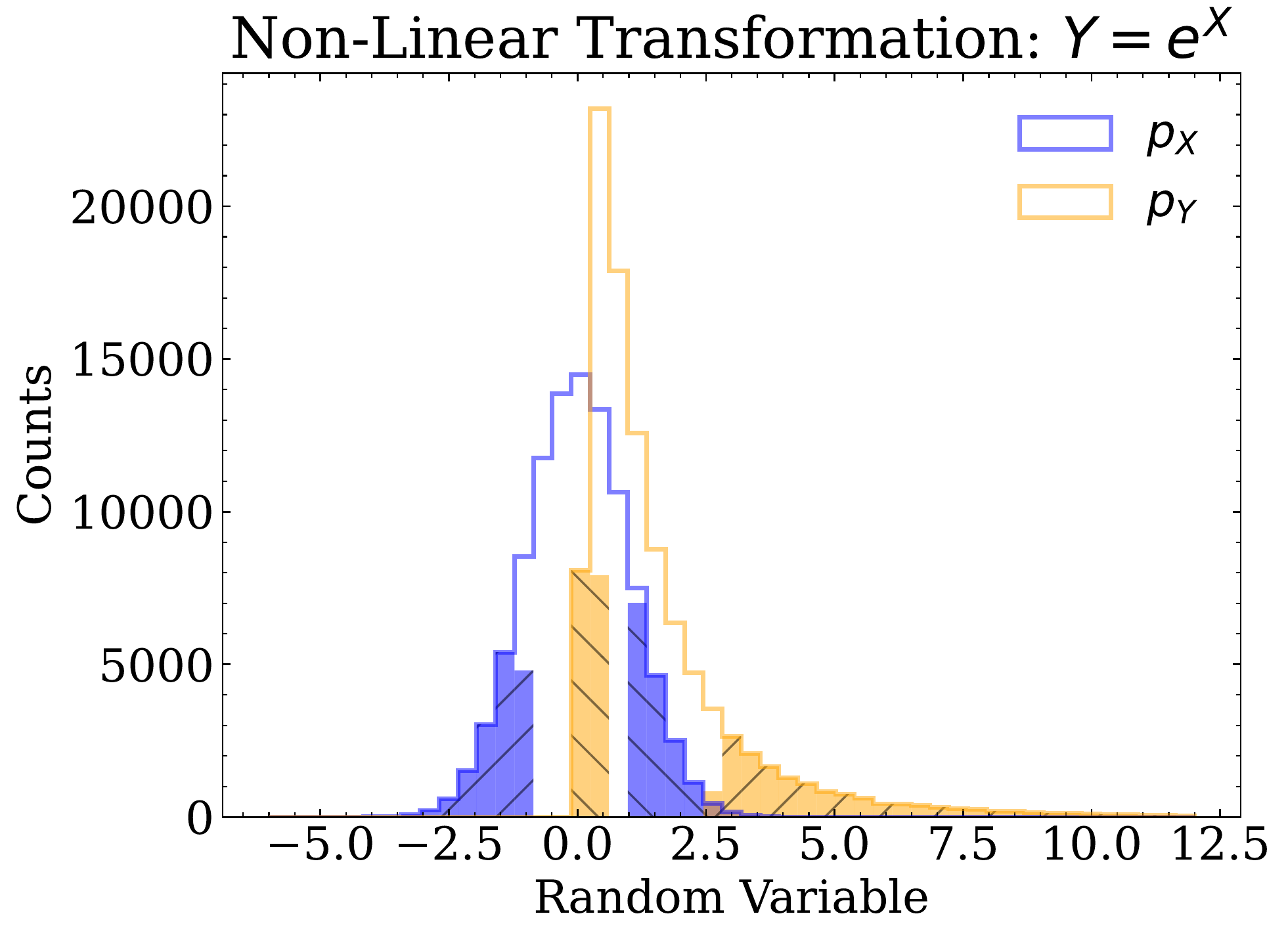}}\\
    \caption{Histograms for a random normal variable $X$ with a (a) linear transform $Y = 2X + 5$, (b) non-linear transform $Y = e^{-X}$. The shaded regions mark where $|X| > 1$. Variables that originate from low-density values of $X$ are hatched (``/" for $X < 1$ and ``\textbackslash" for $X > 1$). For the linear transformation, low-density values of $X$ map to low-density values of $Y$. For the non-linear transformation, however, the low-density values originating from $X > 1$ are mapped to high-density values of $Y$.}
    \label{fig:transforms}
\end{figure}

Suppose that we have initial coordinates $X\in\mathbb{R}^N$ and coordinate transformation $Y=f(X)$, where $f$ is an invertible and differentiable function\footnote{Much of the discussion also still applies if this is not true everywhere, but the bookkeeping becomes significantly more complex, so we focus on this case.}.  If a point in phase space $x$ has probability density $p_X$, then the corresponding point $y=f(x)$ has probability density:

\begin{align}
\label{eq:change}
    p_Y(y)=p_X(f^{-1}(y))\left|\frac{d}{dy}f^{-1}(y)\right|\,,
\end{align}
where the last term is the Jacobian determinant of $f^{-1}$ evaluated at $y$.  If $f$ is a linear transformation, then the Jacobian determinant is independent of $x$. This means that if we order events by density, then the ordering is unchanged. As an example, consider the linear function $y = ax + b$. By the above equation, we have $p_Y(y)=p_X(\frac{y-b}{a})\left|\frac{1}{a}\right|$. Coordinate changes of these types produce a simple shift and rescaling of the probability distribution $p_X$, as shown in Fig.~\ref{fig:transforms}(a).  Note that this includes standardization where the mean is subtracted and then the data are divided by the standard deviation.

In contrast, if $f$ is non-linear, then the Jacobian determinant can depend on $x$. As an example, the non-linear function $y=e^{-x}$ yields the probability density $p_Y(y) = p_X(-\ln(y))\left|\frac{1}{y}\right|$, so the Jacobian determinant is still a function of $y$ and therefore also of $x$. Since the Jacobian determinant is non-constant, this choice of coordinate transformation can dramatically affect the density-ranked order of events, as shown in Fig.~\ref{fig:transforms}: low-density values of $X$ are mapped to high-density values of $Y$.

One popular anomaly detection protocol would be to take events that are `rare' in an absolute sense: $p_X(x) < c$ for some threshold $c$. If $c$ is fixed, then the events selected would change under coordinate transformations due to the Jacobian factor in Eq.~\ref{eq:change}.  An alternative protocol that is more robust (but still sensitive) to coordinate transformations would consider `rare' in a relative sense so that the Jacobian factors cancel.  In particular, instead of comparing densities to an absolute threshold, we could compare the density of one event to the density of other events.  A protocol in this direction would be to take a fraction $q$ of the `rarest' events. 

For example, in one dimension, this corresponds to using a threshold $c$ given by the $q$ quantile of the density.  
Symbolically, the quantile in $X$ for a one-dimensional random variable are given by:\footnote{It may be useful to consider both the highest and lowest quantiles, although for ordering by anomaly score, presumably only the most anomalous events should be considered (not the least). }
\begin{align}\label{eq:quantile}
    q &= \int_{c}^\infty p_X(x) dx\,.
    \end{align}
    Since $c$ is now defined by an integral over a density and not a bare density, one may hope that it is more robust to coordinate transformations.  Ideally, if we compute the threshold $c'$ after transforming into $y=f(x)$, we would have $f(c)=c'$.  In reality:
\begin{align}
    q&=\int_{c'}^\infty p_Y(y) dy\\\nonumber
    &=\int_{c'}^\infty\label{eq:transquantile}
    p_X(f^{-1}(y))\left|\frac{d}{dy}f^{-1}(y)\right| dy\\
    &=\int_{f^{-1}(c')}^{f^{-1}(\infty)} p_X(x)\left|\frac{df}{dx}\right|^{-1} \frac{df}{dx}\,dx\/.
\end{align}
The features $X$ over which the quantile is computed could be the original observables or one could first map to the anomaly score and consider the most anomalous events.
If the Jacobian in Eq.~\ref{eq:transquantile} is non-negative, then the two penultimate terms cancel and $f^{-1}(c')=c$, so the same events are selected before and after the coordinate transformation.  However, if the Jacobian takes on negative values, the order of events under $f$ is reversed and then different events can be selected ($f^{-1}(c')\neq c$). For example, if $f(x)=-x$, then the lowest and highest quantiles are completely reversed.  Another extreme example is when $f$ is the Cumulative Distribution Function (CDF).  In this case, $f(X)$ is uniformly distributed between $0$ and $1$ so no point is rarer than any other.

In contrast, likelihood ratio methods are invariant under coordinate transformations because the Jacobian determinant in Eq.~\ref{eq:change} is the same for the target probability density and the reference probability density (and thus drops out in the ratio). This is strictly only true when $f$ is bijective (as assumed above), but it may be approximately true even if this is not the case.  Note that even though likelihood ratios are formally invariant under coordinate transformations, it may be that practical approaches benefit from a judicial choice of coordinates.  For example, observable standardization is often essential in enabling effective ML training.

Equation~\ref{eq:change} is a well-known fact found in textbooks of probability and statistics.  Its connection with anomaly detection was recently made by the machine learning community~\cite{e23121690}. In the following section, we provide an illustrative Gaussian example and then make an explicit connection with HEP, both using the relative threshold protocol.

\section{Numerical Examples}
\label{sec:examples}

\subsection{Analytic Case}

\begin{figure}[h!]
    \centering
    \subfloat[]{\includegraphics[width=0.45\textwidth]{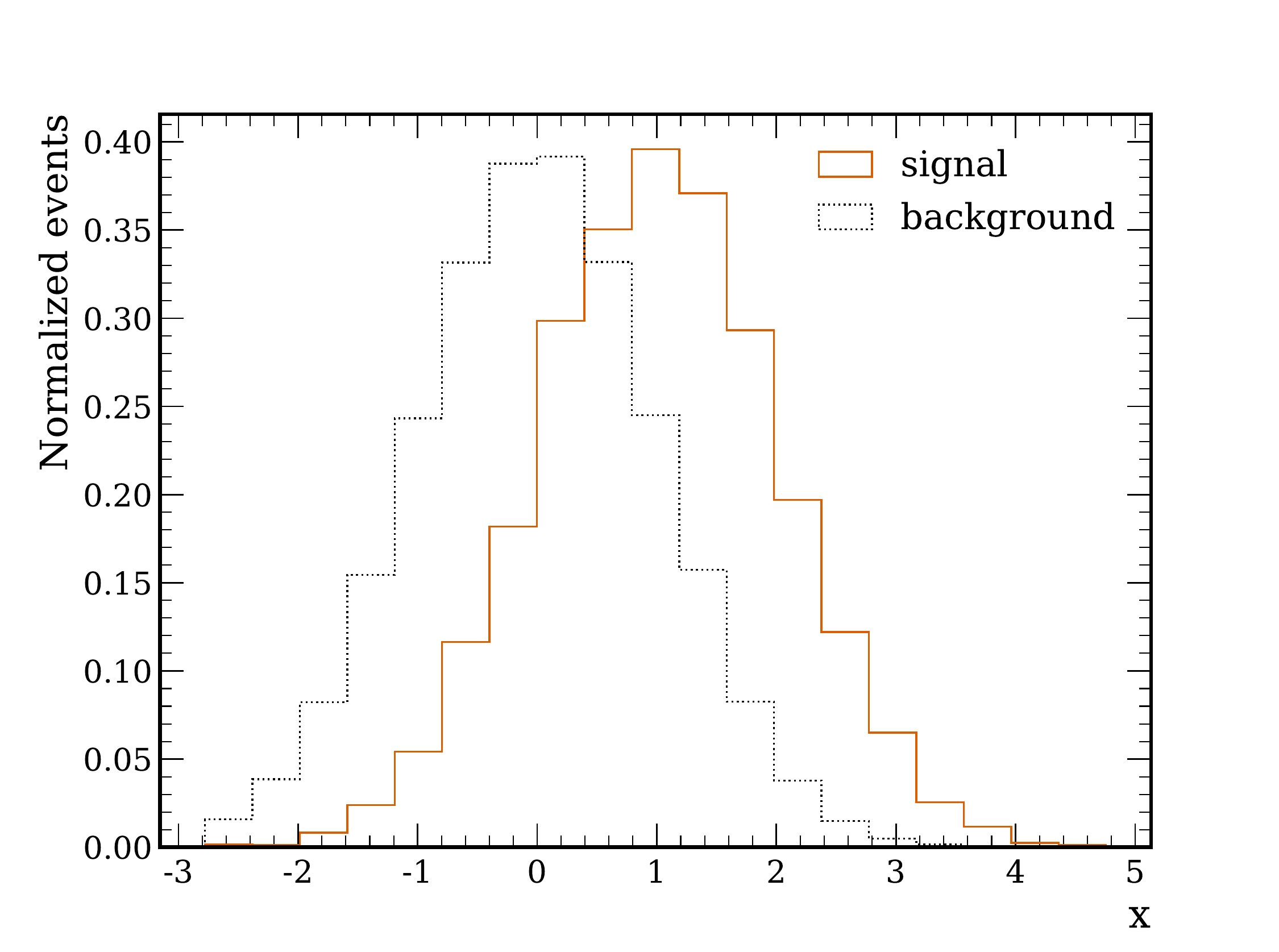}}\\
    \subfloat[]{\includegraphics[width=0.45\textwidth]{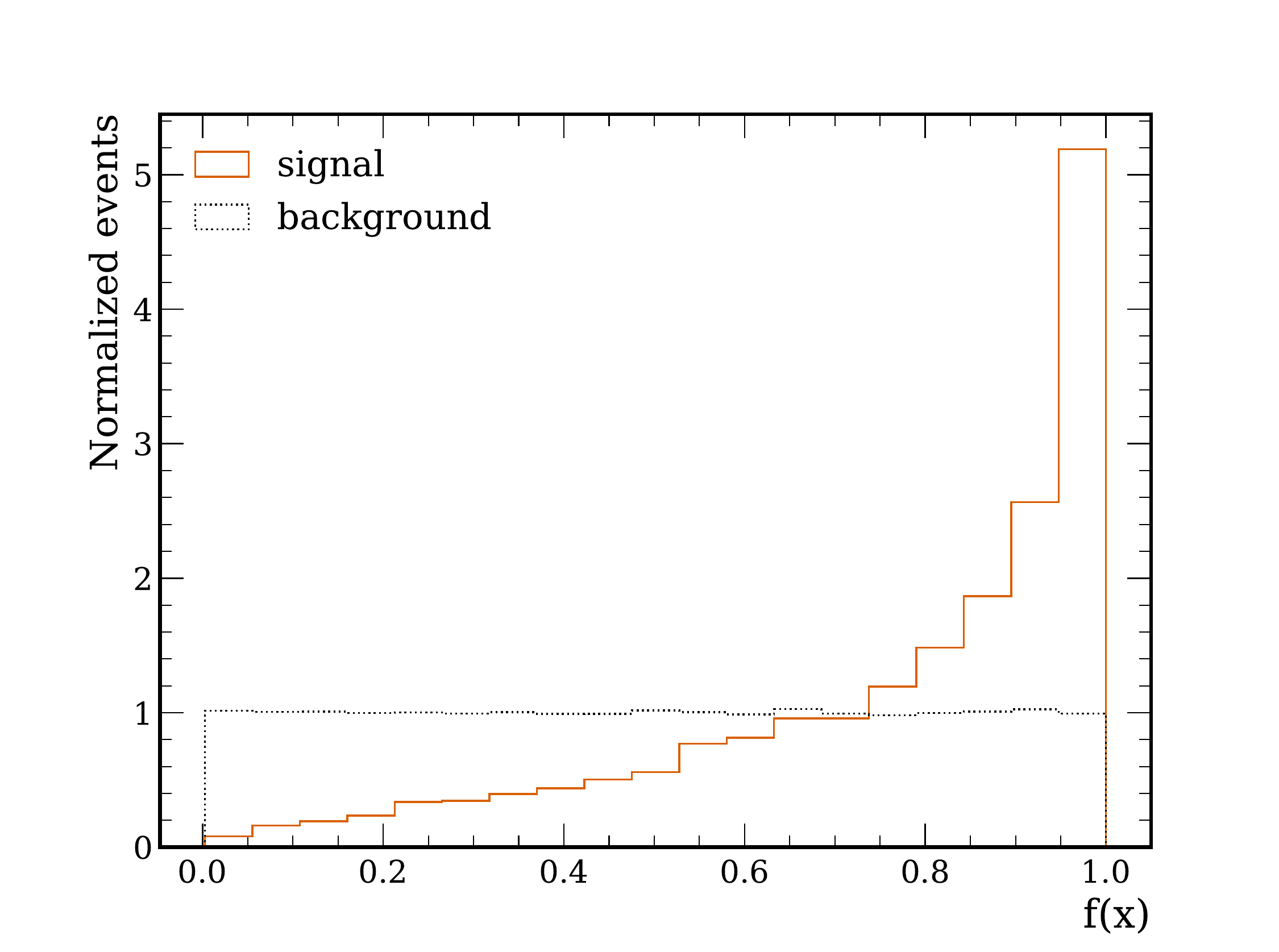}}\\
    \subfloat[]{\includegraphics[width=0.45\textwidth]{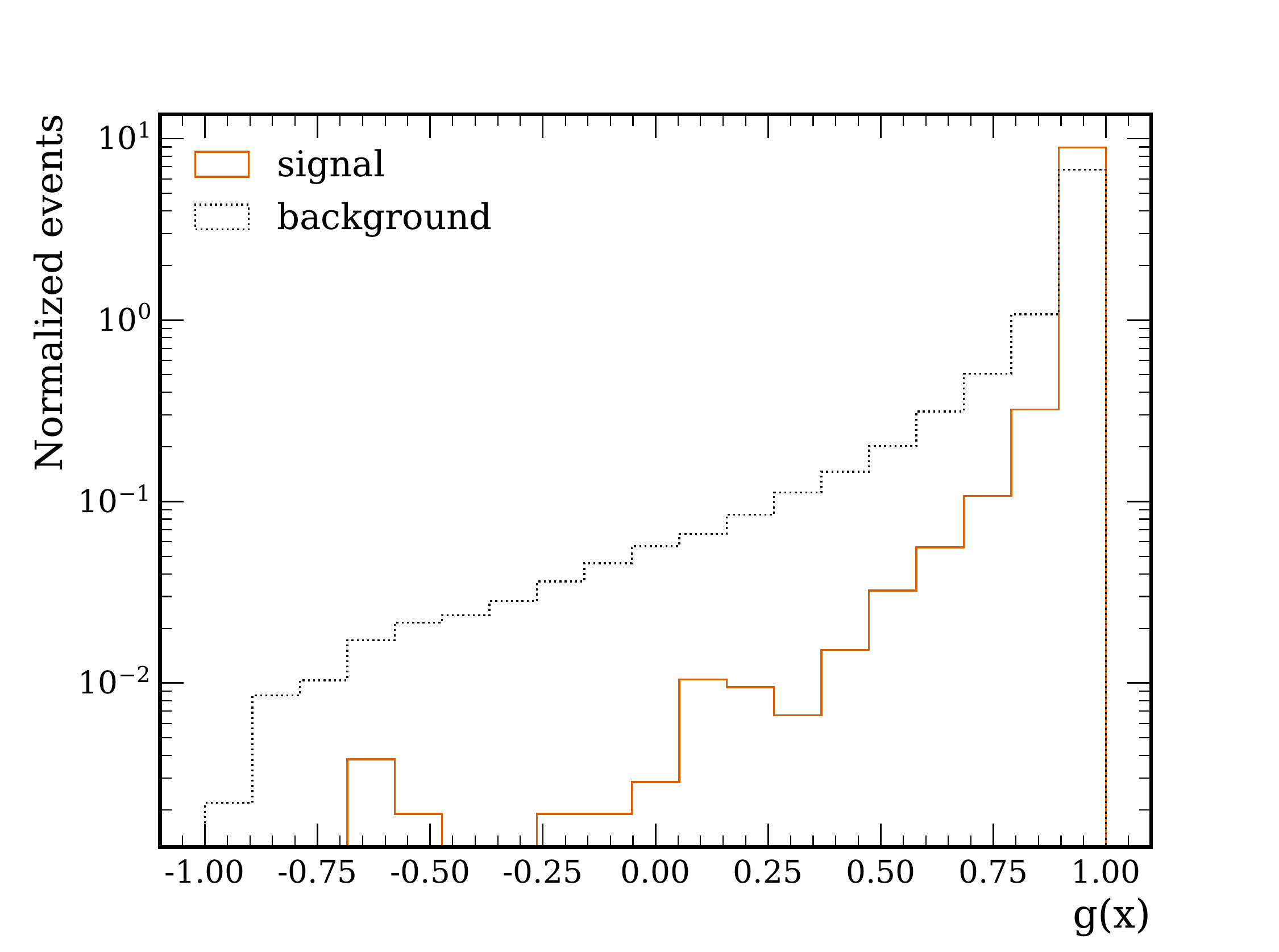}}\\
    \caption{Histograms for the background and signal events in the Gaussian example for (a) the nominal coordinates, (b) after the coordinate transformation $Y = f(X)$ where $f$ is the CDF of a standard normal, and (c) after the coordinate transformation $Y = g(X) = \tanh(X+2)$.}
    \label{fig:gaussian}
\end{figure}

To clearly illustrate the ideas discussed in the previous section, we will construct a simple example to demonstrate a dramatic consequence of this sensitivity to coordinate transformations.  Let $X_b\sim \mathcal{N}(0,1)$ represent a set of background observables, and let $X_s\sim\mathcal{N}(1,1)$ represent a set of signal observables.  This scenario is illustrated in Fig.~\ref{fig:gaussian}(a). A density estimation-based search for anomalies would consist of learning the density of the background $p_{X_b}$, then making a cut where the density is low. This would designate the two tails of $X_b$ as rare, and a search for anomalies would then successfully pick up the signal events $X_s$ overlapping with the right-tail phase space of the background.

Now, suppose that instead of the variables $X_b$ and $X_s$, we used $Y_{b} = f(X_{b})$ and $Y_{s} = f(X_{s})$, where $f$ is the CDF of a standard normal random variable. This scenario is illustrated in Fig.~\ref{fig:gaussian}(b).  In this case, $Y_b$ (but not $Y_s$) would be distributed uniformly from $0$ to $1$. A density estimation-based anomaly detection search would then fail: while the signal is mapped to high values under the transformation $Y_s = f(X_s)$, there are no anomalous (i.e. low-density) regions of the background variable $Y_b$ that would be identified and probed for signal.
    
One could imagine even less optimal transformations that produce high background densities where there are high signal densities and low background densities where there are low (or zero) signal densities. One such scenario is illustrated in Fig.~\ref{fig:gaussian}(c) for the transformation $Y_{b,s} = g(X_{b,s}) = \tanh(X_{b,s}+2)$. Anomaly detection through density estimation would fail for such a transformation of variables due to the background distribution aligning closely with the signal distribution.

We also illustrate the impact of a change of coordinates when popular anomaly detection algorithms are used to identify the anomalies. We train an Autoencoder, a Normalizing Flow, and a weakly-supervised model based on the Classification Without Labels (\textsc{CWoLa}) paradigm~\cite{Metodiev:2017vrx,Collins:2018epr,Collins:2019jip}. The dataset before the change of coordinates consists of two-dimensional distributions of background  $X_b\sim \mathcal{N}(0,1)$ and signal  $X_s\sim \mathcal{N}(1,1)$, with each dimension independent and identically distributed. The two-dimensional dataset is used to ensure the bottleneck layer of the Autoencoder is lower dimensional than the input. The two functions used are the same ones introduced previously: $f$, i.e. the CDF of a standard normal random variable,  and $g(x) = \tanh(x+2)$.

\begin{figure}[ht]
\centering
\includegraphics[width=0.4\textwidth]{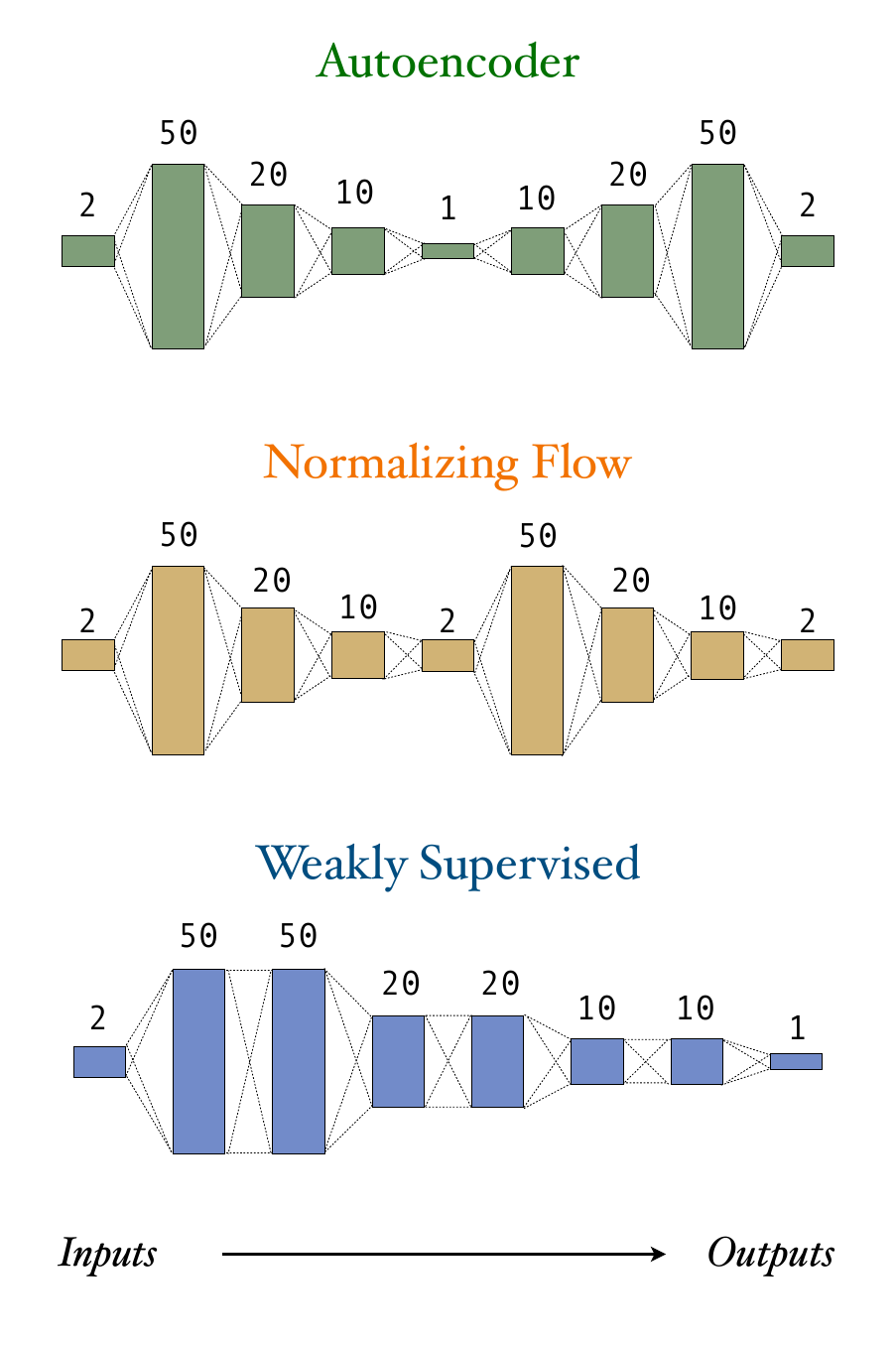}
\caption{Network architectures used to implement the anomaly detection methods.}
\label{fig:arch}
\end{figure}

The Autoencoder compresses the two-dimensional data into a one-dimensional latent space using fully-connected layers of sizes (50, 20, 10) and \textsc{ReLU} activation functions before the bottleneck layer of size 1. The decoder is simply the mirrored version of the encoder architecture. Only background events are used during training, and the anomaly score is then defined by the reconstruction loss. The Normalizing Flow is built using a continuous Normalizing Flow~\cite{grathwohl2018ffjord} with a backbone neural network defined by two stacked fully-connected models with layer sizes (50, 20, 10) and $\tanh$ activation. The background-only density is estimated with anomaly score defined as minus the probability density of a single event. Finally, the weakly-supervised model based on \textsc{CWoLa} is trained using a classifier consisting of six fully-connected layers of sizes (50, 50, 20, 20, 10, 10) and the \textsc{ReLU} activation function. The model is  trained to separate a reference sample of background only events from a mixed sample of signal plus background events, with signal events representing 10\% of the overall dataset size. The anomaly score is taken as the ratio $h(x)/(1-h(x))$, where $h(x)$ is the classifier output after a sigmoid activation function. A summary of the anomaly detection methods and the anomaly scores is given in Tab.~\ref{tab:algorithms} with the different model architectures shown in Fig.~\ref{fig:arch}. All methods are implemented using \textsc{TensorFlow}~\cite{tensorflow2015-whitepaper} and \textsc{Adam}~\cite{adam} optimizer with learning rate of 0.001 for 500 epochs or until the validation loss, assessed using an independent dataset, does not improve for 10 consecutive epochs. 

\begin{table}[ht]
    \centering
	\small
    \caption{Choice of anomaly detection methods and anomaly scores used in this work.}
    \label{tab:algorithms}
	\begin{tabular}{lcc}
        Algorithm & Anomaly score\\
        \hline
        Autoencoder & $(g(f(x))-x)^2$\\
        Normalizing flow & -$p_b(x)$\\
        Weakly-supervised & $h(x)/(1-h(x))$

	\end{tabular}
\end{table}

\begin{figure}[ht]
\centering
\includegraphics[width=0.49\textwidth]{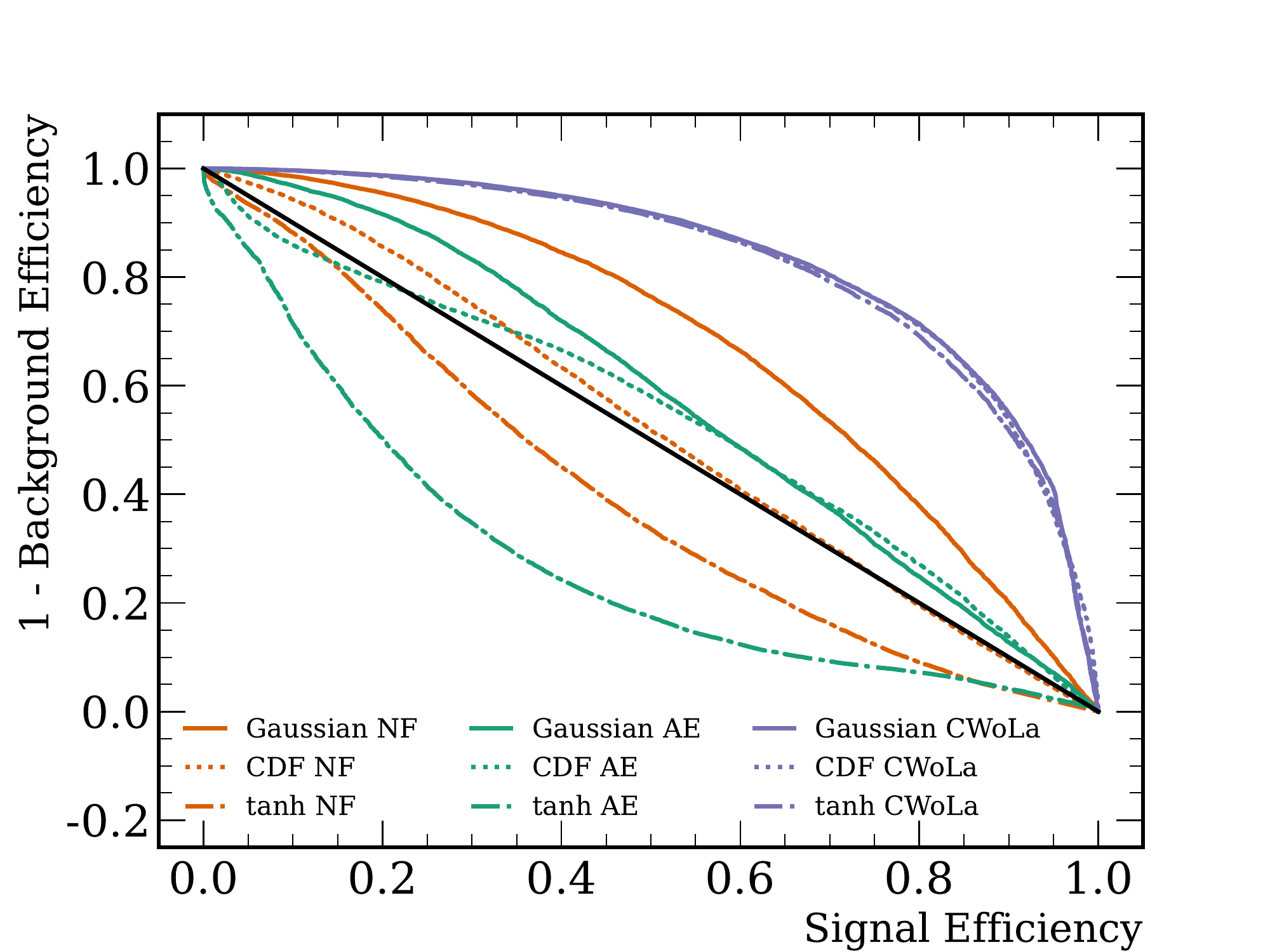}
\caption{Receiver operating characteristic (ROC) curve for different anomaly detection algorithms trained using the Gaussian dataset in the original coordinate system (Gaussian) and after the transformations $f(x) = \Phi(x)$ (CDF) and $f(x) = \tanh(x+2)$ (tanh). The black line denotes the expected ROC curve for a random algorithm.}
\label{fig:roc_gaus}
\end{figure}

We evaluate the performance for each algorithm using the Receiver Operating Characteristic (ROC) curve for signal and background events, as shown in Fig.~\ref{fig:roc_gaus}.

Both the Autoencoder and Normalizing Flow show reasonable performance in the Gaussian example, but both fail to identify the anomaly after the CDF change of coordinates and even end up systematically removing signal events after the hyperbolic tangent transformation. The weakly-supervised algorithm, on the other hand, identifies the anomaly and shows the same performance for all choices of coordinate systems. 

\subsection{LHC Olympics Case}

The examples in the previous section were contrived in order to demonstrate the most extreme cases.  This section uses realistic HEP observables where the impact is not as dramatic, but the effects of coordinate transformations are still non-negligible.

The dataset used here was originally developed for the LHC Olympics~\cite{Kasieczka:2021xcg} and is briefly described in the following.  The background process is dijets and the signal is $W^{\prime} \to X (\to q q) Y(\to q q) $ with $m_{W^{\prime}} = 3.5$~TeV, $m_X = 500$~GeV, and $m_Y = 100$~GeV.  All events are generated using \textsc{Pythia8}~\cite{Pythia2} and \textsc{Delphes3.4.1}~\cite{delphes1, delphes2, delpes3}. The jets are clustered using \textsc{FastJet}~\cite{fastjetManual} with the anti-$k_T$ algorithm~\cite{anti-kt} using $R=1$. Finally, all events are required to have at least one jet with $p_T > 1.2$~TeV.

Some important discriminating features in the LHC Olympics dataset are the masses of the leading and subleading jets.  In particular, the masses of the leading ($m_1$) and subleading ($m_2$) jets should approximately correspond to the masses of the $X$ and $Y$ particles for the signal.  Since the masses have a large kinematic range, they are often preprocessed by taking the natural logarithm, $m\mapsto \log(m/\text{TeV})$ (henceforth, the units are suppressed).  Other natural examples include the $n$-subjettiness observables $\tau_1$ and $\tau_2$~\cite{nsubjetiness1,nsubjetiness2}.  These observables quantify the extent to which a jet is more consistent with having one or two prongs.  The variable $\tau_1$ captures similar properties of the jet radiation pattern as the jet mass.  A researcher aiming to pre-process as minimally as possible might attempt to do anomaly detection with $(\tau_1,\tau_2)$ directly, while someone wanting to use standard pre-processing might use instead $(\tau_1,\tau_2/\tau_1)$.  The $n$-subjettiness ratio $\tau_{21}=\tau_2/\tau_1$ is one of the most widely used taggers for identifying two-prong substructure.  This is characteristic of Lorentz-boosted $W/Z$ boson decays, but it is also the case for our BSM particles $X$ and $Y$.  We show results for $m$ and $\log(m)$, but we found similar, although less dramatic, results for $n$-subjettiness.

If $(m_1,m_2)$ is described by probability density $p$, then the transformed coordinates are described by density  $\tilde{p}(\log(m_1),\log(m_2))=p(m_1,m_2)m_1m_2$.  This shows that the ordering by anomaly score can be reversed depending on the relative sizes of $p$, $m_1$ and $m_2$.

\begin{figure}[t]
    \centering
    \includegraphics[width=0.45\textwidth]{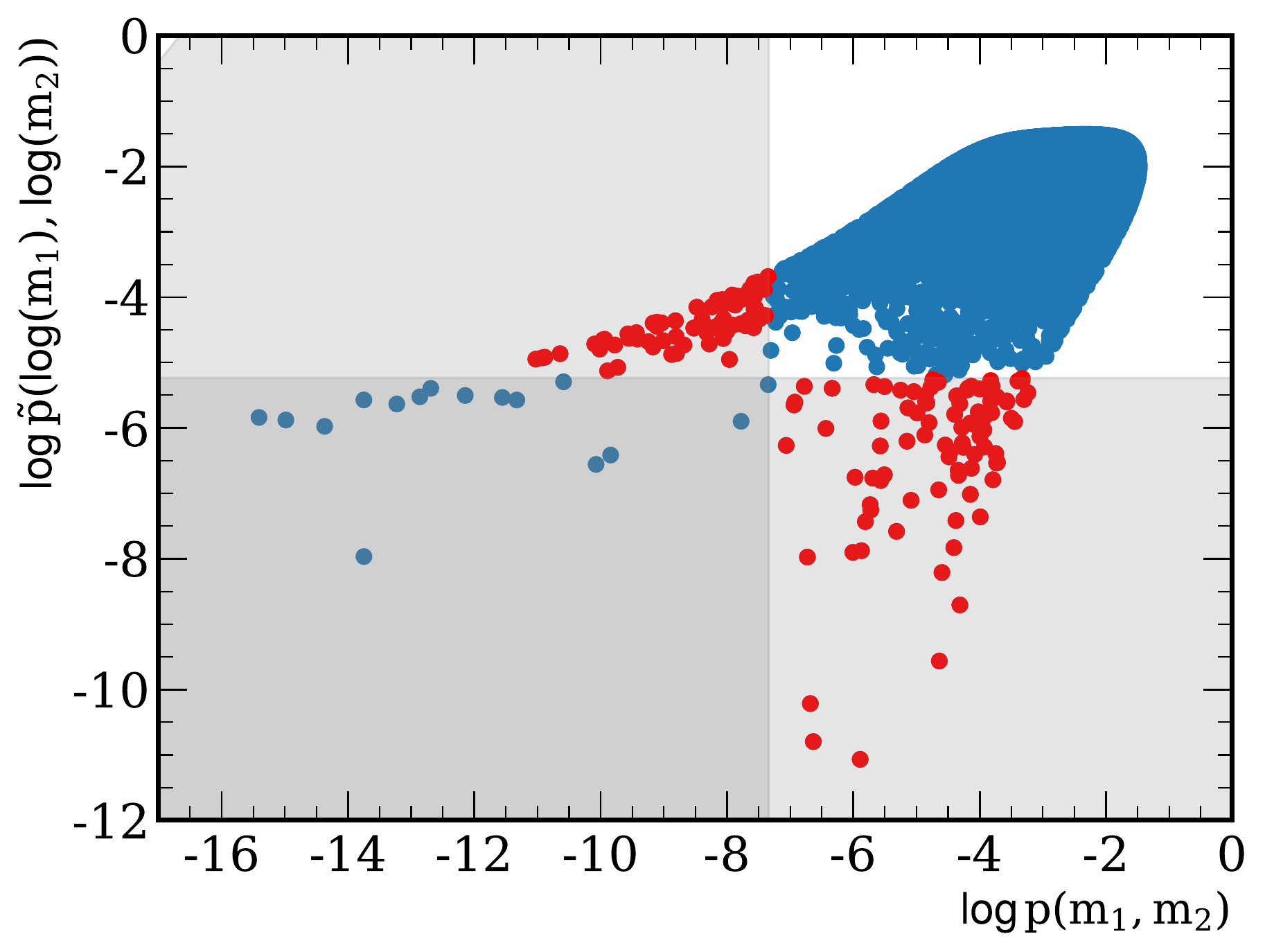}
    \caption{A comparison of the two probability densities for the same events.  The shaded regions and the red dots indicate the 1\% most anomalous events. }
    \label{fig:nsub}
\end{figure}

Unlike in the Gaussian case, for the LHC Olympics dataset we do not know the probability densities analytically and so we can only estimate the densities numerically.  A comparison of the selected anomalies in the background-only case with a NF are presented in Fig.~\ref{fig:nsub}. %
The shaded regions in Fig.~\ref{fig:nsub} indicate the selected anomalies using a 1\% criteria.  Interestingly, the two selections agree on only about 20\% of events.  This means that \textbf{even though we have the same events and the same input features, we  have different anomaly selections depending on the coordinates we use to represent the events.}  %

\begin{figure}[ht]
\centering
\subfloat[]{\includegraphics[width=0.45\textwidth]{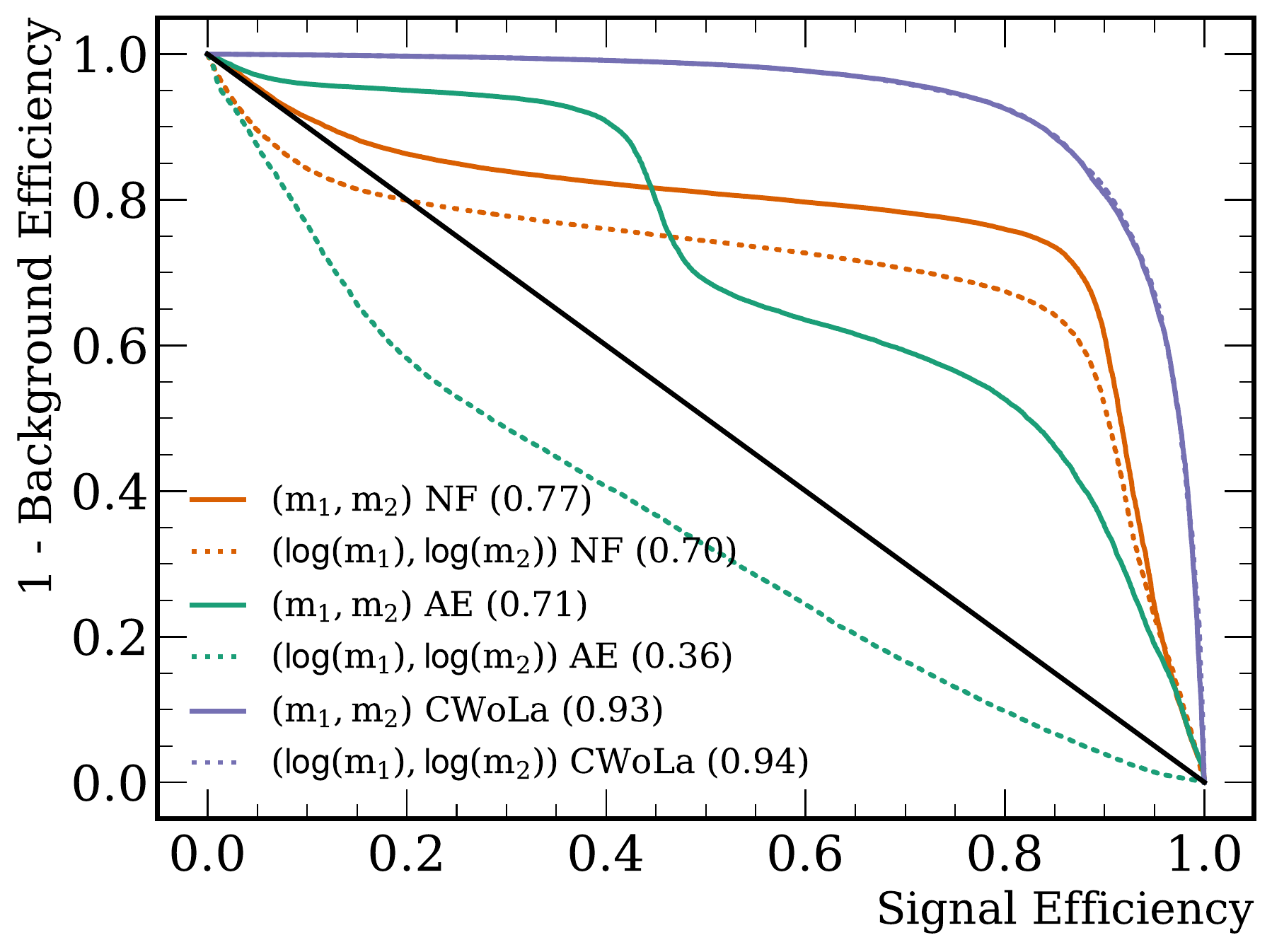}}\\
\subfloat[]{\includegraphics[width=0.45\textwidth]{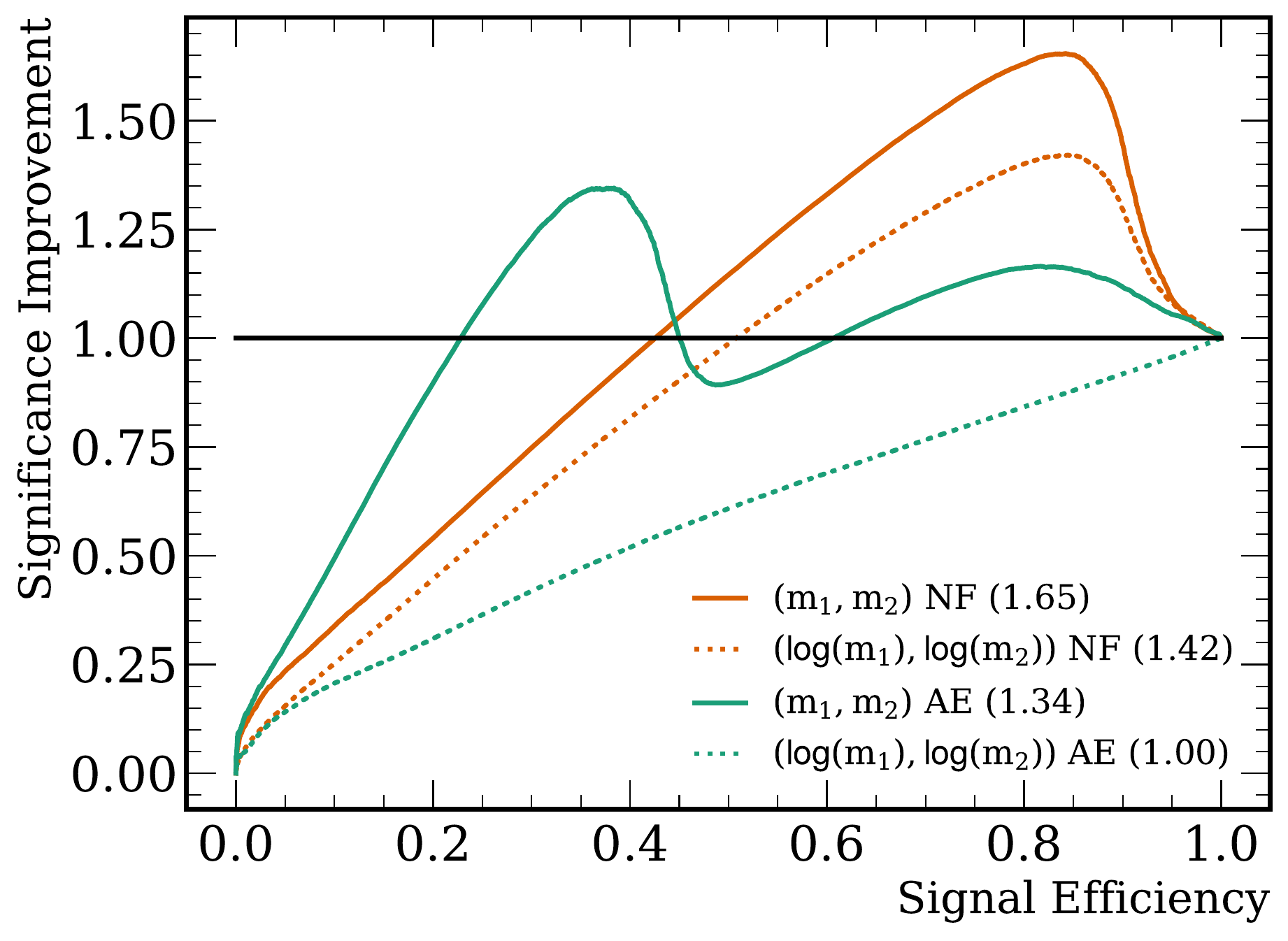}}
\caption{(a) Receiver Operating Characteristic (ROC) curves for the three anomaly detection algorithms evaluated using the LHC Olympics dataset for different choices of inputs.  The black line denotes the expected ROC curve for a random algorithm. The number in parentheses represents the area under the curve.  (b) Same as (a), but instead of the background efficiency, the dependent variable is the Significance Improvement Characteristic (SIC) defined as the signal efficiency divided by the square root of the background efficiency.}
\label{fig:roc_lhco}
\end{figure}

Analogously to the previous section, we compare anomaly detection strategies in Fig.~\ref{fig:roc_lhco}. We employ the same neural network models and hyperparameters as the ones used in the Gaussian example.  Once again, the performance of the weakly-supervised training is independent from the choice of coordinates, while all other algorithms show differences in performance based on the initial choice of coordinates.  The AE and NF have a similar performance, reinforcing the claim that the approaches are targeting similar regions of phase space.  However, the change in performance after the coordinate transformation is more pronounced for the AE, which may have other contributions aside from the indirect density estimation. 

\section{Conclusions and Outlook}
\label{sec:conclusions}

In this paper, we have described the sensitivity of anomaly detection approaches to coordinate transformations.  We have connected BSM hypotheses with ML strategies in order to make explicit what is being assumed and when coordinate transformations are potentially important.  While it is not new, we have highlighted the coordinate sensitivity of unsupervised approaches (targeting `rare' events).  There is no optimal set of coordinates \emph{a priori}, but for a given signal hypothesis, some set of coordinates will be optimal for a particular unsupervised learning algorithm.  This does not mean that we should not use unsupervised algorithms -- on the contrary, these approaches provide valuable complementarity to other less-than-supervised methods.  However, we should be cautious about optimal claims, and it seems wise to explore multiple coordinate systems when determining the sensitivity.  While weakly-supervised approaches are formally coordinate-independent, it could be that in practice some set of coordinates enables more efficient learning.  These and other practical issues are critically important to explore as anomaly detection proposals become physics results in the near future.

\section*{Code Availability}

The code for this paper can be found at \url{https://github.com/ViniciusMikuni/Rareisnotuniversal}.

\section*{Acknowledgments}

We are grateful to the participants, speakers, and co-organizers of the PhyStat-Anomaly workshop (\url{https://indico.cern.ch/event/1138933/}) for many useful discussions before, at, and after the meeting. We additionally thank the participants of Hammers and Nails 2022 for useful discussions, especially Kyle Cranmer. We also thank Sascha Caron, Bob Cousins, and Louis Lyons for feedback on the manuscript.  RM, VM, BN, and MP are supported by the U.S. Department of Energy (DOE), Office of Science under contract DE-AC02-05CH11231. This research used resources of the National Energy Research Scientific Computing Center, a DOE Office of Science User Facility supported by the Office of Science of the U.S. Department of Energy under Contract No. DE-AC02-05CH11231 using NERSC award HEP-ERCAP0021099.  This material is based upon work supported by the National Science Foundation Graduate Research Fellowship Program under Grant No. DGE 2146752. Any opinions, findings, and conclusions or recommendations expressed in this material are those of the author(s) and do not necessarily reflect the views of the National Science Foundation. The work of DS was supported by DOE grant DOE-SC0010008.
GK acknowledges support by the Deutsche Forschungsgemeinschaft (DFG, German Re\-search Foundation) under Germany’s Excellence Strategy – EXC 2121  ``Quantum Universe" – 390833306.  


\bibliography{HEPML,other}
\bibliographystyle{apsrev4-1}

\end{document}